# Delivery of nanosecond laser pulses by multi-mode anti-resonant hollow core fiber at 1 µm wavelength

**MENG ZHAO,**[1,2] **FEI YU,**[1,3,6] **DAKUN WU,**[3] **XINYUE ZHU,**[1,7] **SI CHEN,**[1] **MENG WANG,**[1] **MINZHE LIU,**[4] **KUN ZHAO,**[4] **RUIZHAN ZHAI,**[4] **ZHONGQING JIA,**[4] AND **JONATHAN KNIGHT**[5]

[1]*Key Laboratory of Materials for High Power Laser, Shanghai Institute of Optics and Fine Mechanics, Chinese Academy of Sciences, Shanghai 201800, China*
[2]*Center of Materials Science and Optoelectronics Engineering, University of Chinese Academy of Sciences, Beijing 100049, China*
[3]*Hangzhou Institute for Advanced Study, University of Chinese Academy of Sciences, Hangzhou 310024, China*
[4]*Laser Institute, Qilu University of Technology (Shandong Academy of Sciences), Qingdao, Shandong, 266000, China*
[5]*Centre for Photonics and Photonic Materials, Department of Physics, University of Bath, Claverton Down, Bath, BA2 7AY, UK*
[6]*yufei@siom.ac.cn*
[7]*zhuxinyue@siom.ac.cn*

**Abstract:** In this paper we explore the application of low-loss multimode anti-resonant hollow-core fiber (MM-AR-HCF) [1] in the delivery of nanosecond laser pulses at 1 µm wavelength. MM-AR-HCF of large core offers a rich content of low-loss higher-order modes which plays a key role in the efficient coupling and transmission of high-power laser of degraded beam quality. In the experiment, laser pulses of an average pulse energy of 21.8 mJ with 14.6 ns pulse width (corresponding a peak power of 1.49 MW) are transmitted through MM-AR-HCF of 9.8 m length without damaging. Up to 94 % coupling efficiency is achieved where the incident laser beam suffers a degraded beam quality with $M_x^2$ and $M_y^2$ of 2.18 and 1.99 respectively. Laser-induced damage threshold (LIDT) of MM-AR-HCF measures 22.6 mJ for 94 % coupling efficiency, which is 7 times higher than that for multimode silica optical fiber with a core diameter of 200 µm.

## 1. Introduction

High-power lasers have found increasing applications in many industrial fields, such as laser cutting, welding and additive manufacturing [1-4]. Optical fiber based laser power by is known for its light weight, flexibility, robustness and low cost, which set free many constraints of using bulky laser sources in the practical industrial applications and scientific researches.

The newly emerging anti-resonant hollow-core fiber (AR-HCF) confines the light in the large hollow core providing a nearly free-space-like guidance for the laser delivery. AR-HCFs exhibit lower optical nonlinearity, lower dispersion and higher laser induced damage threshold (LIDT) than solid-core silica optical fibers, which have been demonstrated in the high-power and high-field laser delivery in the spectral region from the deep ultraviolet to the mid-infrared [5-7]. It is noted that most AR-HCF designs reported present the quasi-single mode guidance for higher-order modes (HOMs) are usually left with higher losses by being coupled with cladding modes under phase matching conditions [8-11]. Therefore, nearly all demonstrated high-power AR-HCF delivery experiments necessitate a fairly good beam quality of laser source with $M^2$ less than 1.5 usually [7, 12].

To cater for the delivery of high-power laser beam of degraded beam quality, the concept of multimode anti-resonant hollow-core fiber (MM-AR-HCF) is proposed. In 2019, Winter and colleagues fabricated a MM-AR-HCF with an inner diameter of 164 µm and an outer diameter of 360 um [13]. By enlarging the diameter ratio of core over cladding, the phase-matching condition can be tuned so as to bring down the attenuations of certain HOMs of low rank. However, such a large diameter of core significantly increases bending loss, making it difficult to apply in the laser delivery. In 2022, Shere et. al. proposed two designs of MM-AR-HCF for high-power industrial laser delivery and simulated the modal properties sysmatically [14]. Recently, Wu et. al. designed and fabricated low-loss MM-AR-HCFs with 18 fan-shaped resonators in the cladding for near and mid-infrared spectral regions, and properties of multimode guidance was characterized by $S^2$ method [15].

Micro-structured hollow-core fibers including photonic-bandgap hollow-core fibers (PBG-HCFs) and Kagome hollow-core fibers (Kagome-HCF) have demonstrated advantages of delivery of Q-switched nanosecond laser pulses over traditional solid-core fibers in terms of higher LIDT. In 2004, Shepard et al. firstly applied a 7-cell photonic-bandgap hollow-core fiber (PBG-HCF) to successfully transmit nanosecond pulses from Nd: YAG with pulse energy of up to 0.37 mJ (65 ns pulse width) [16]. Later, a 19-cell PBG-HCF was demonstrated the delivery of higher pulse energy of 1.025 mJ (10 ns pulse width) with an optimized coupling efficiency of nearly 82 % [17]. In 2013, Wang et al. applied Kagome-type hollow-core fibers in the delivery of Nd: YAG laser pulses and brought up the transmitted pulse energy up to 4 mJ (9 ns pulse width) [18]. In the same year, a quasi-single-mode AR-HCF was firstly explored in the transmission of nanosecond laser pulses where the laser source presented a good beam quality with $M^2$ of 1.2 [19]. More details are referred to Table 1.

In this paper, we demonstrate the damage-free delivery of nanosecond pulses with energy of 21.8 mJ and a corresponding peak power of 1.49 MW. At the output of MM-AR-HCF, $M^2$ of transmitted beam is found to be reduce from about 2 to 1.7. LIDT at the incident end of MM-AR-HCF measures an average pulse energy of 22.6 mJ that is 7 times higher than that of a commercial multimode silica fiber with a large core of 200 μm in diameter.

Table 1. Delivery of Nd: YAG nanosecond laser pulses by micro-structured hollow-core fibers

| Years | Fiber | Beam Quality ($M^2$) | Coupling Efficiency (%) | Maximum incident pulse energy (mJ) | Pulsed width (ns) | Peak power (MW) | Ref |
|---|---|---|---|---|---|---|---|
| 2004 | 7-cell PBG-HCF | 1.2 | / | 0.37 | 65 | 0.006 | [16] |
| 2007 | 19 cell PBG-HCF | 1.5 | 82 % | 1.025 | 10 | 0.10 | [17] |
| 2011 | Kagome HC-PCF | / | 30 % | 1.3 | 7 | 0.19 | [20] |
| 2013 | HC-NCF | 1.2 | 92 % | 1.1 | 60 | 0.02 | [19] |
| 2013 | 1-cell Kagome fiber | / | 39 % | 4 | 9 | 0.44 | [18] |
| 2013 | 7-cell Kagome fiber | / | 89 % | 10 | 9 | 1.11 | [18] |
| 2023 | Multi-mode AR-HCF | 2.18-X 1.99-Y | 94 % | 21.8 | 14.6 | 1.6 | This work |

## 2. Properties of MM-AR-HCF

### 2.1 Attenuation of MM-AR-HCF

The homemade MM-AR-HCF was fabricated by the stack-and-draw technique with a core diameter of 66 μm and an outer diameter of 193 μm. The cladding is composed of 18 fan-shape capillaries. And the average core wall thickness is about 350 nm where the corresponding first resonant wavelength is calculated at 740 nm approximately [1]. The measured attenuation at 1064 nm is 0.044 dB/m by the cutback from 56.9 to 30 meters. More detailed on the characterization of bend loss and mode content are found in [15].

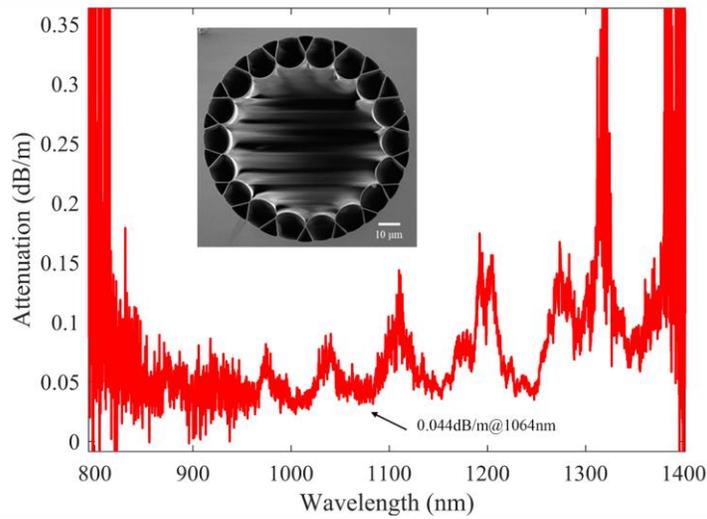

Fig.1 The measured attenuation of MM-AR-HCF by the cut back method. Inset: Scanned electronic microscopy picture of MM-AR-HCF. The core diameter is 66 μm and core wall thickness is 350 nm.

## 2.2 Numerical aperture of MM-AR-HCF

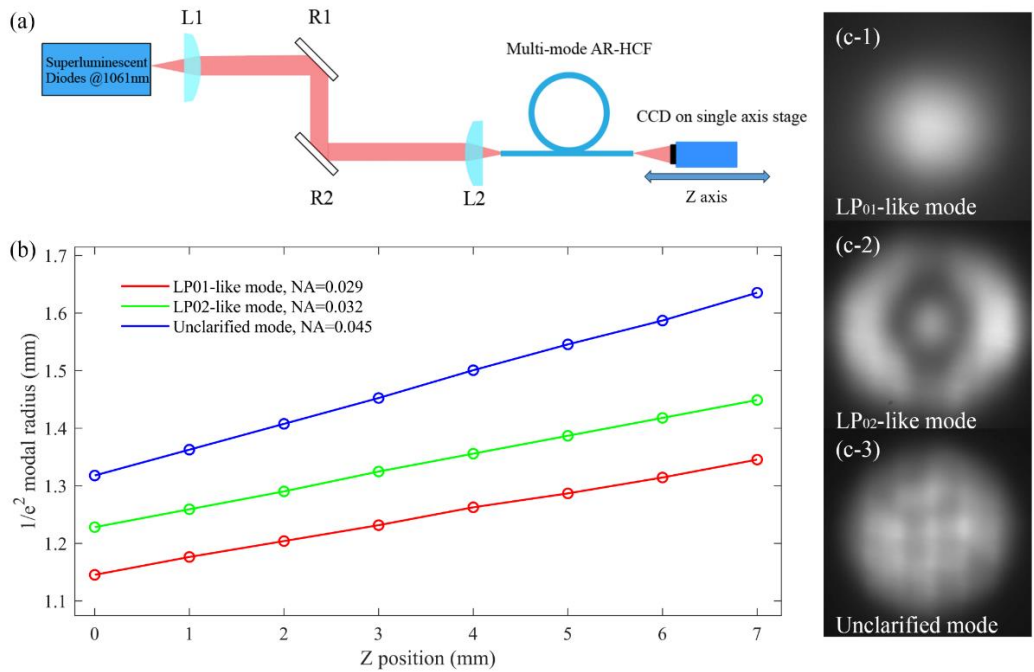

Fig.2 (a) Schematic of experiment to measure the NA of MM-AR-HCF. L1 and L2 are microscopic objective lenses; R1 and R2 are silver mirrors (with average reflection R>97%@1064 nm); A CCD camera is mounted on a single-axis translation stage to record the far-field patterns of emission from MM-AR-HCF along Z axis. (b) Modal diameters of far-filed patterns as functions of z position. 8 meters of MM-AR-HCF is used. Corresponding near-field images at the output end of MM-AR-HCF which are: (c-1) $LP_{01}$ -like mode, (c-2) $LP_{02}$ -like mode, (c-3) unclarified mode.

Numerical aperture (NA) of MM-AR-HCF is characterized by scanning the far field of mode emitted at the fiber end as shown in Fig.2 (a). A superluminescent diode (Thorlabs, S5FC1050P) is used as the light source with the central wavelength at 1061 nm and full width at half maxima of 50 nm. For different coupling conditions, $LP_{01}$-like mode, $LP_{02}$ -like mode and some unclarified field distribution are excited and their NAs are measured as 0.029, 0.032 and 0.045 respectively.

## 3. Experimental study of MM-AR-HCF delivery of nanosecond laser pulses

### 3.1 Experiment setup

Figure 4 illustrates the schematic of laser delivery experiment by using MM-AR-HCF. A Q-switched Nd: YAG nanosecond pulse laser source (Continuum Surelite SL I-10) is used. The laser output is a linearly polarized elliptical beam with long and short axes of 5.147 mm and 3.312 mm at the exit of laser respectively. The temporal width of pulse is 14.6 ns and the repetition rate 10 Hz. In the experiment, the laser power coupled in MM-AR-HCF can be continuously adjusted by rotating the half-wave plate (HWP) before the polarizer.

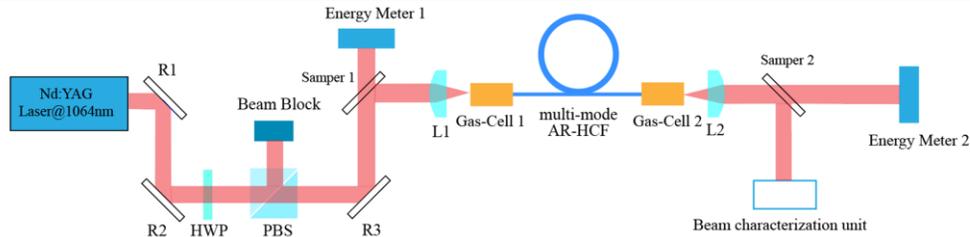

Fig.3 Setup of nanosecond pulse laser delivery experiment. R1, R2 and R3 are Nd: YAG Laser Line Mirrors with reflection > 99.5 %; L1 and L2 are plano-convex lenses with $f_1$=75 mm, $f_2$=50 mm; Samper 1 and 2 are beam splitters with 90:10 ratio; HWP: half wave plate; PBS: polarization beam splitter. Two ends of MM-AR-HCF are sealed in homemade gas-cells for vacuuming.

A 9.8 m MM-AR-HCF is used in the laser delivery which is loosely rewound on an iron plate with a bend radius of 30 cm. To prevent the damage from the air ionization excited by laser pulses, both MM-AR-HCF ends are sealed in the homemade gas cells and vacuumed. Both internal pressure in the gas cells are maintained around 1.02 mbar in the experiment.

### 3.2 Characterization of coupling and transmission efficiencies

Figure 4 (a) shows the laser transmission through the 9.8 m long MM-AR-HCF as function of the incident pulse energy. In the experiment, the output power of laser source is sampled and monitored to calibrate the incident and output powers of MM-AR-HCF. As the incident power rises, the transmission efficiency stays around 85 %, corresponding a coupling efficiency of 94 % by subtracting the measured attenuation of 0.044 dB/m at 1064 nm wavelength.

Before damage, the maximum single transmitted pulse energy is recorded as 23.4 mJ with a peak power of 1.6 MW accordingly. The average of pulse energy over 30 seconds is 21.8 mJ in this case. Figure 5 (b) shows the temporal measurement of laser transmission over 20 minutes for an average incident pulse energy of 18 mJ.

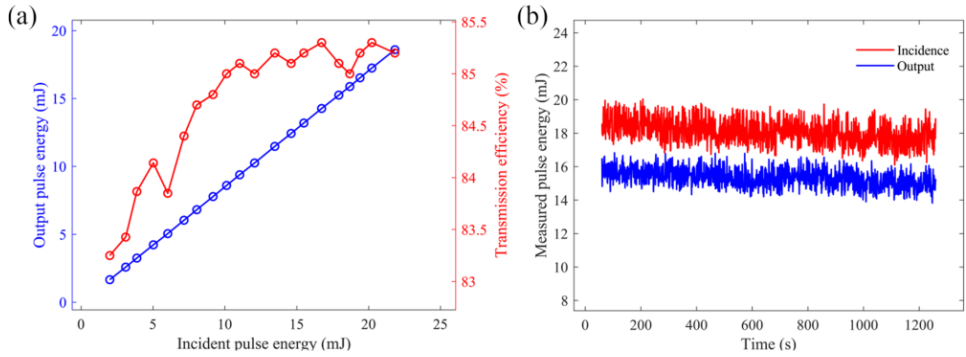

Fig.4 (a) The measured averaged transmitted pulse energy over 9.8 m MM-AR-HCF length and calculated transmission efficiency as function of incident pulse energy. (b) The temporal measurement of laser transmission over 20 minutes for an averaged incident pulse energy of about 18mJ.

### 3.3 Characterization of transmitted pulse widths and spectra

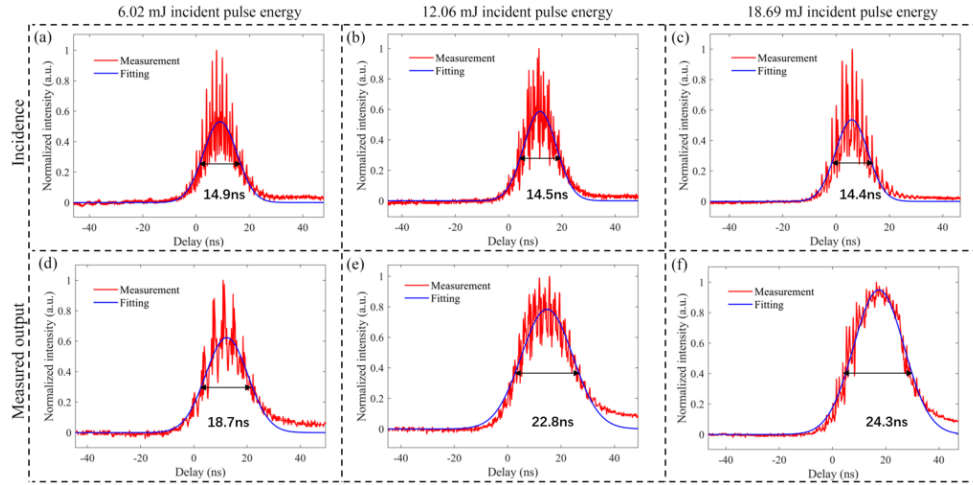

Fig.5 Normalized temporal measurement of different incident and transmitted pulse profiles. Incident pulses with energies of (a)6.02 mJ, (b)12.06 mJ, (c)18.69 mJ and corresponding output with energy of (d) 5.05 mJ, (e)10.25 mJ, and (f)15.88 mJ. The temporal broadening for high power incidence is attributed to the light propagation in the silica cladding.

The measured temporal profiles of transmitted pulses are presented in Fig. 6 for different incident pulse energies of 6.02 mJ, 12.06 mJ and 18.69 mJ respectively. An InGaAs detector (Thorlabs, DET08CFC/M) and oscilloscope (Keysight, DSOX6002A) are used in the measurement. The full-width at half maxima of the Gaussian fitting of pulse profile is calculated as the pulse width. The pulse width of delivered laser pulses increases from about 14 ns to 24.3 ns for higher incident power. According to the $S^2$ measurement in [15], the contribution of higher-order modes to the pulse broadening is no more than 40 ps for a propagation length of 9.8 m. We attribute the measured temporal broadening to light propagation in the silica cladding. The group delay of light propagating in the silica from in the air core is about 14.7 assuming the group refractive indices of silica and core mode are 1.45 and 1.

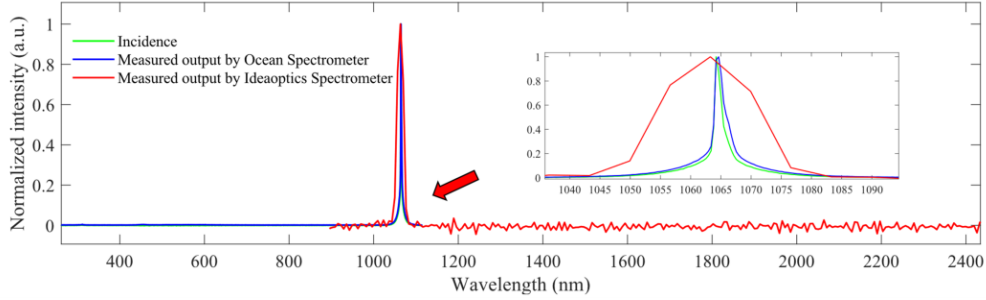

Fig.6 Normalized optical spectra of laser pulses before and after transmission in the 9.8 m long MM-AR-HCF when the incident pulse energy is 18.69 mJ. The output spectrum is slightly broadened over the incidence. Inset: zoom-in of spectrum at 1064 nm wavelength. The different widths between the red line and blue or green line are due to different resolutions of two spectrometers in use.

The spectra of laser beam before and after transmission in the 9.8 m MM-AR-HCF are measured by spectrometers for an incident pulse energy of 18.69 mJ. The incident laser beam is measured by Ocean spectrometer (Maya 1000Pro), and the output by both Ocean and Ideaoptics Spectrometer (NIR25S) to scan the entire visible and near-infrared spectral regions. Compared to the incidence, the output spectrum at 1064nm is slightly broadened by approximately 0.5 nm. In the vacuumed length of MM-AR-HCF, we attribute the spectral broadening to the possible self-phase modulation from a part of laser power transmitted in the silica cladding.

### 3.4 Characterization of beam quality before and after transmission

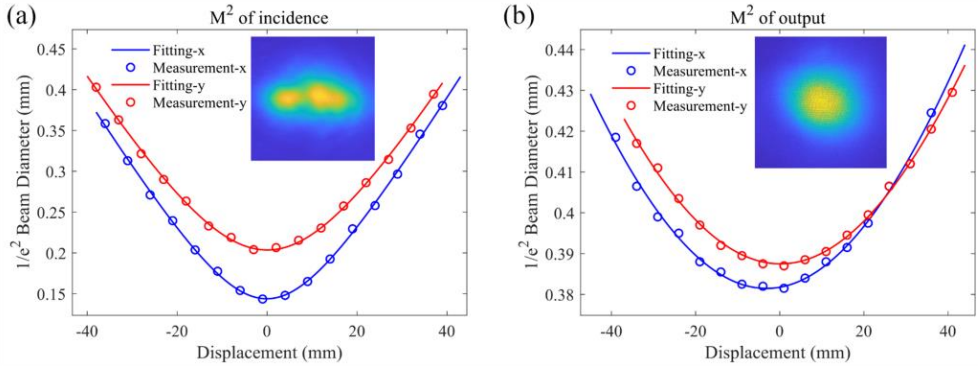

Fig.7 The input(a) and output(b) beam diameters of far-field patterns from the output of MM-AR-HCF along the z direction (dots). Insets: typical far-filed patterns imaged by the CCD.

The beam profiles before and after transmission in MM-AR-HCF are characterized by $M^2$ measurement using the knife-edge method. As Fig. 7 shows, the output of laser source has an elliptical profile of beam with $M_x^2$ and $M_y^2$ as 2.18 and 1.99 respectively. The beam quality at the output end of MM-AR-HCF is improved with $M_x^2$ and $M_y^2$ reduced to 1.70 and 1.75.

### 3.5 Comparison of laser-induced damages of MM-AR-HCF and large-core multimode silica optical fiber

When deviating from the optimized coupling condition, laser-induced damage (LID) to the end of MM-AR-HCF is found even for an incident pulse energy as small as 3.3 mJ as shown in Fig 9. For 94 % coupling efficiency, LIDT of MM-AR-HCF reaches 23.4 mJ and beyond.

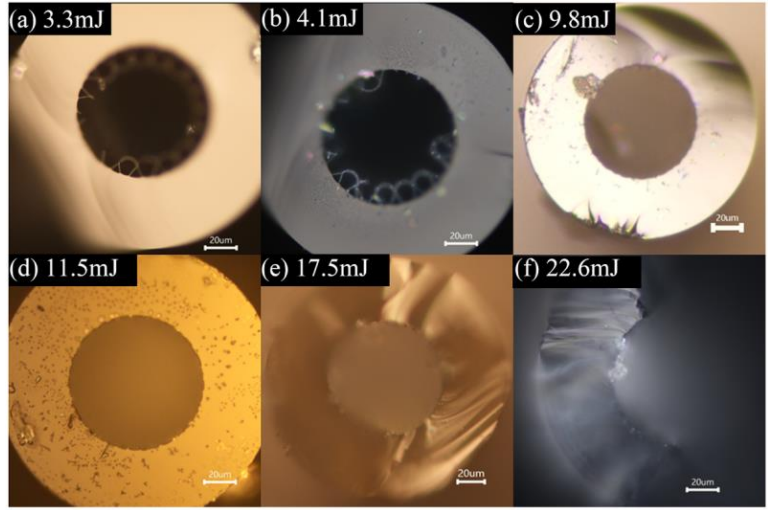

Fig.8 Microscopic images of damaged MM-AR-HCF ends for incident pulse energies of (a) 3.3 mJ, (b) 4.1 mJ, (c) 9.8 mJ, (d) 11.5 mJ, (e) 17.5 mJ, (f) 22.6 mJ when the coupling condition deviates from the optimal.

In comparison, we maintain the same laser delivery experiment setup but replace MM-AR-HCF with a large-core multimode silica optical fiber (MMF). The MMF (Xinrui, SIH200 22A) has a core diameter of 200±2 μm, a cladding diameter of 220±5 μm, and a numerical aperture of 0.22±0.02. As the incident pulse energy reaches 3 mJ, some point damages and cracks are to be observed at the end of MMF. It is noted that such damages are often accompanied with sparks in the air near the fiber end. When the incident pulse energy rises to 4.8 mJ, the end of MMF would be completely destroyed.

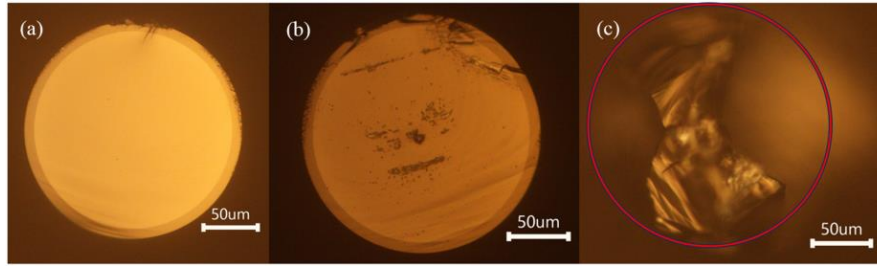

Fig.9 Microscopic images of MMF ends: (a) before damage (b) and (c) for an incident energy of 3 mJ and 4.8 mJ respectively. The red circle in (c) indicates the fiber end region before damage.

## 4. Conclusion

In this paper, a homemade MM-AR-HCF is experimentally demonstrated in the degraded laser beam delivery of nanosecond pulses at 1 μm wavelength. In spite of incident beam quality with $M^2$ up to 2, about 94 % coupling efficiency is still achieved. And after propagating through the 9.8 m MM-AR-HCF, the laser beam has $M^2$ reduced to 1.7. LIDT of MM-AR-HCF is found around 7 times higher than that for a MMF with a core diameter of 200 μm. Our experiment implies promising potential of AR-HCFs in high-power laser delivery of various application requirement.

**Funding.** Chinese Academy of Sciences (ZDBS-LY JSC020); National Natural Science Foundation of China (61935002, 62075200, 62127815); Key R&D Program of Shandong Province (2021CXGC010202).

**Acknowledgments.** We would like to thank Mr. Jinhu Zheng and Mr. Henan Shen for assistance in the fiber drawing.

**Disclosures.** The authors declare no conflicts of interest.

**Data availability.** The data used to produce the plots within this work is found in Ref. [21] .